\documentclass{article}
\def\be{\begin{equation}}
\def\eea{\end{eqnarray}}
\def\bea{\begin{eqnarray}}
\def\ee{\end{equation}}

\usepackage[psamsfonts]{amssymb}
\usepackage{amsmath}
\usepackage{cite}
\author{M. Alimohammadi$^{1}$ \footnote{alimohmd@ut.ac.ir} and M.
Khorrami$^2$ \footnote{mamwad@mailaps.org}
\\ $^1$ {\small Department of Physics, University of Tehran,}
\\ {\small North Karegar Ave., Tehran, Iran.}
\\ $^2$ {\small Department of Physics, Alzahra University, Tehran
19938-91167, Iran.} }
\title{Large-$N$ limit of the two-dimensinal Yang-Mills theory on
surfaces with boundaries}
\date{}
\begin{document}
\maketitle
\begin{abstract}
\noindent The large-$N$ limit of the two-dimensional
U$(N)$ Yang-Mills theory on an arbitrary orientable compact surface
with boundaries is studied. It is shown that if the holonomies of the
gauge field on boundaries are near the identity, then the critical behavior
of the system is the same as that of an orientable surface without boundaries
with the same genus but with a modified area. The diffenece between this
effective area and the real area of the surface is obtained and shown to be a
function of the boundary conditions (holonomies) only. A similar result
is shown to hold for the group SU$(N)$ and other simple groups.
\end{abstract}
\section{Introduction}
In recent years, the two-dimensional Yang-Mills theory (YM$_2$)
has been studied by many authors \cite{1,2,3,4,5,6,7}. It is an
important integrable model which can shed light on some basic
features of QCD$_4$. Also, there exists an equivalence between
YM$_2$ and the string theory. It was shown that the coefficients
of the $1/N$ expansion of the partition function of SU$(N)$ YM$_2$
are determined by a sum over maps from a two-dimensional surface
onto the two-dimensional target space.

The partition function and the expectation values of the Wilson loops of
YM$_2$ have been calculated in lattice- \cite{1,8} and
continuum-formulations \cite{4,9,10,19}. All of these quantities are described
as summations over the irreducible representations of the corresponding gauge
group. In general, it is not possible to perform these summations explicitly.
For large gauge groups, however, these summations may be dominated by some
specific representations, and it can be possible to perform the
summations explicitly. There are other physical reasons as well (for example
the relation between large-$N$ YM$_2$ and the string theory), that the
study of the YM$_2$ for large groups is important.

In \cite{11}, the large-$N$ limit of the U$(N)$ YM$_2$ on a sphere
was studied. There it was shown that the above mentioned summation is
replaced by a (functional) integration over the continuous parameters
of the Young tableaux corresponding to the representation. Then the
saddle-point approximation singles out a so-called classical representation,
which dominates the integration. In this way, it was shown that the free energy
of the U$(N)$ YM$_2$ on a sphere with the surface area $A<A_{\mathrm{c}}=\pi^2$
has a logarithmic behavior \cite{11}. In \cite{12}, the free energy was
calculated for areas $A>\pi^2$, from which it was shown that the YM$_2$ on
a sphere has a third-order phase transition at the critical area
$A_{\mathrm{c}}=\pi^2$, like the well known Gross-Witten-Wadia
phase transition for the lattice two dimensional multicolour gauge theory
\cite{13,14}. The phase structure of the large-$N$ YM$_2$, generalized YM$_2$,
and nonlocal YM$_2$ on a sphere were further discussed in \cite{20,27,37,39}.

It can be easily seen that for surfaces with no boundaries, only
the genus-zero surface (i.e. the sphere) has a nontrivial
saddle-point approximation, and all other surfaces have trivial
large-$N$ behavior \cite{11}.

For surfaces with boundaries, the situation is much more involved.
In these cases, for each boundary, the character of the holonomy
of the gauge filed corresponding to that boundary appears in the
expression of the partition function. These characters complicate
the saddle-point equation. In \cite{15}, the large-$N$ properties
of YM$_2$ on cylinder and on zero-area vertex manifold (a sphere
with three holes) have been studied. If we denote the $j$'th
boundary by $C_j$, then each boundary condition is specified by
the conjugacy class of the holonomy matrix
$U_j=$Pexp$\oint_{C_j}{\rm d}x^\mu\;A_\mu(x)$. So, the boundary
condition corresponding to $C_j$ is fixed by the eigenvalues of
$U_j$. These eigenvalues are unimodular, so that each of them is
specified by a real number $\theta$ in $[-\pi,\pi]$. In the
large-$N$ limit, the eigenvalues of these matrices become
continuous and one can denote the set of these eigenvalues
(corresponding to the $j$'th boundary) by an eigenvalue density
$\sigma_j(\theta),\theta\in [-\pi,\pi]$.

In the case of cylinder where only two boundaries $U_1$ and $U_2$
exist, it has been shown that the free energy of YM$_2$ satisfies
a Hamilton-Jacobi equation with a Hamiltonian describing a fluid
of a negative pressure. The time coordinate of the system is the
area of the cylinder between one end and a loop ($0\leq t \leq
A$), and its position coordinate is $\theta$, with two boundary
conditions $\sigma (\theta )|_{t=0}=\sigma_1 (\theta )$ and
$\sigma (\theta )|_{t=A}=\sigma_2 (\theta )$. In this way it was
shown that the classical Young tableau density $\rho_c$ at
large-$N$, satisfies $\pi\rho_c[-\pi \sigma_0(\theta )]=\theta$
where  $\sigma_0(\theta )$ is $\sigma (\theta ,t)$ at a time
(area) $t$ at which the fluid is at rest. The presence of a phase
transition is then related to the existence of a gap in the eigenvalue
density $\sigma (\theta )$ \cite{15}. Note that by this method,
only the existence of the phase transition is proved and no
explicit expression can be obtained for critical area. Only for a
disc, $\sigma_2(\theta )=\delta (\theta )$, an explicit expression
for critical area $A_c$ has been given, using the
Itzykson-Zuber integral at large-$N$ limit \cite{16}. Similar
calculation were performed for the large-$N$ generalized YM$_2$
and nonlocal YM$_2$, in \cite{33} and \cite{49} respectively.

For a vertex manifold, and other spheres with more than two
holes, the above Hamilton-Jacobi description does not work since
the time coordinate can not be defined in a vertex (3-way) point.
Therefore, in \cite{15} only a selection rule has been
obtained for a zero area vertex.

In this paper we want to study the large-$N$ behaviour of YM$_2$
on an orientable genus-$g$ surface with $n$ boundaries
($\Sigma_{g,n}$). As explained above, to investigate the partition
function and hence the possible phase transition of these systems
is difficult. So we restrict ourselves to the cases in which the
boundary holonomies $U_j$s are very close to identity. In this
case, one can expand the character $\chi_R(U)$, where $R$ denotes
the irreducible representations of U$(N)$, around $U=I$. It will
be shown that the critical behavior of YM$_2$ on $\Sigma_{g,n}$
with area $A(\Sigma_{g,n})$ is the same as a genus-$g$ surface
with no boundary ($\Sigma_{g,0}$), but with the area
$A(\Sigma_{g,0})=A(\Sigma_{g,n})+V(U_1,\dots,U_n)$. Therefore, it
is seen that in the large-$N$ limit, the phase structure of YM$_2$
on $\Sigma_{g>0,n}$ is trivial, while YM$_2$ on $\Sigma_{0,n}$
exhibits a third-order phase transition, as long as all boundary
holonomies are close to identity.

The plan of the paper is as following. In section 2, the expansion of
$\chi_R(U)$ with $U\in\mathrm{U}(N)$, is obtained in terms of the
eigenvalues $s_1,\dots,s_N$ of $U$, up to second order in $(s_j-1)$'s.
In section 3, the large-$N$ limit of the partition function of U$(N)$ YM$_2$
on $\Sigma_{g,n}$, and its critical behavior is obtained. In the special case
$(g=0,n=1)$, where the surface is a disc, it is shown that our result
coincides with the only known one, obtained by Gross and Matytsin \cite{15}.
Finally in section 4, a similar result for simple gauge groups is obtained
by a different method.
\section{The U$(N)$ characters}
The partition function of a YM$_2$ on a
two-dimensional surface $\Sigma_{g,n}$ is \cite{4}
\begin{equation}\label{z}
 Z_{g,n}(U_1,\dots,U_n;A)=\sum_Rd_R^{2-2g-n}\chi_R(U_1)
\cdots\chi_R(U_n)\,e^{-\frac{A}{2N}C_{2\,R}}.
\end{equation}
$C_{2\,R}$ is the second Casimir of the group in the
representation $R$, $A$ is the surface area, and the factor
$N^{-1}$ (as the coupling for the gauge group U$(N)$ or SU$(N)$)
has been inserted to give the system a nontrivial saddle-point
expansion. As it can be seen from (\ref{z}), corresponding to each
boundary a factor $\chi_R(U_i)/d_R$ appears in the expression for
the partition function. Let us expand this factor around $U\approx
I$ for the group U$(N)$.

The representation $R$ of the gauge group U$(N)$ is characterized
by $N$ integers $l_1$ to $l_N$, satisfying
\begin{equation}
+\infty >l_1>l_2>\cdots >l_N>-\infty .
\end{equation}
The group element $U$ has $N$
eigenvalues $s_1=e^{i\theta_1}$ to $s_N=e^{i\theta_N}$. The
character $\chi_R(U)$ is then
\begin{equation}\label{ch}
  \chi_R(U)=\frac{\det\left\{e^{i l_j\theta_k}\right\}}
          {\mathrm{van}(s_1,\dots ,s_N)},
\end{equation}
where $\mathrm{van}(s_1,\dots ,s_N)$ is the van der Monde
determinant
\begin{align}
\mathrm{van}(s_1,\dots ,s_N):=&
\begin{vmatrix}
s_1^{N-1} & s_1^{N-2} &\cdots &1 \\
s_2^{N-1} & s_2^{N-2} &\cdots &1 \\
\vdots &\vdots & & \vdots \\
  s_N^{N-1} & s_N^{N-2} &\cdots &1
\end{vmatrix},\nonumber\\
=& \prod_{i<j} (s_i-s_j).
\end{align}
From (\ref{ch}) it is seen that if $s_0$ is a phase and $U=s_0\,U'$, then
\begin{equation}\label{5}
\chi_R(U)=s_0^{l_1+\cdots+l_N-[N(N-1)/2]}\,\chi_R(U').
\end{equation}
Any member of U$(N)$ can be decomposed as a product of a phase and a member
of SU$(N)$. The above result then shows the relation between the character of any
element of U$(N)$ and the character of the corresponding element of SU$(N)$.
(One can take $U'$ to be in SU$(N)$, which means that the product of its
eigenvalues is equal to one.)

Denoting the numerator of (\ref{ch}) by $F(\mathbf{s},\mathbf{l})$:
\begin{align}\label{f}
 F(\mathbf{s},\mathbf{l}):=&{\det\left\{e^{i l_j\theta_k}\right\}},\nonumber\\
=&\begin{vmatrix}
s_1^{l_1} & s_2^{l_1} &\cdots &s_N^{l_1} \\
s_1^{l_2} & s_2^{l_2} &\cdots &s_N^{l_2} \\
\vdots &\vdots & & \vdots \\
  s_1^{l_N} & s_2^{l_N} &\cdots &s_N^{l_N}
\end{vmatrix},
\end{align}
it is seen that it has roots at $l_i=l_j$ and $s_i=s_j$, so it is
proportional to $\mathrm{van}(s_1,\cdots ,s_N)$ and
$\mathrm{van}(l_1,\cdots ,l_N)$. Expanding the
remaining part around $\mathbf{s}=(1,\dots,1)=:\mathbf{e}$, it is found that
\begin{align}\label{lr}
F(\mathbf{s},\mathbf{l})=\mathrm{van}(s_1,\dots ,s_N)\,
\mathrm{van}(l_1,\cdots ,l_N)\Bigg\{&E+B\sum_i(s_i-1)+C\sum_i(s_i-1)^2
\nonumber\\& +D\sum_{i<j}(s_i-1)(s_j-1) +
\cdots \Bigg\}.
\end{align}
Defining
 \begin{equation}
 \xi_i:=\ln s_i,
\end{equation}
and
 \begin{equation}
  {\cal O}=\left(\frac{\partial}{\partial\xi_1}\right)^{N-1}
\left(\frac{\partial}{\partial\xi_2}\right)^{N-2}\cdots
\left(\frac{\partial}{\partial\xi_{N-1}}\right),
  \end{equation}
acting by ${\cal O}$ on (\ref{lr}), and putting $\mathbf{s}=\mathbf{e}$,
one arrives at
\begin{equation}
\mathrm{van}(l_1,\dots ,l_N)=E\,\mathrm{van}(N,N-1,\dots ,1)
\,\mathrm{van}(l_1,\dots ,l_N),
 \end{equation}
from which one arrives at
 \begin{equation}
 E=\frac{1}{\mathrm{van}(N,N-1,\dots ,1)}.
\end{equation}

Defining
\begin{equation}
{\cal O}_1:=\left(\frac{\partial}{\partial\xi_1}\right){\cal O},
\end{equation}
applying ${\cal O}_1$ on (\ref {lr}), and putting $\mathbf{s}=\mathbf{e}$,
one arrives at
\begin{equation}\label{13}
\Bigg(\sum_i l_i \Bigg)\mathrm{van}(l_1,\dots,l_N)=
{\rm van}(l_1,\cdots ,l_N)\Bigg(q_N+\frac{N\,B}{A}\Bigg),
\end{equation}
where
\begin{equation}
\frac{\partial^n}{\partial\xi^n}=s^n\left(\frac{\partial}{\partial s}\right)^n +
q_n\,s^{n-1}\left(\frac{\partial}{\partial s}\right)^{n-1} +
p_n\,s^{n-2}\left(\frac{\partial}{\partial s}\right)^{n-2} + \cdots
\end{equation}
has been used in which
\begin{align}
q_n&=\frac{n(n-1)}{2},\nonumber\\
p_n&=\frac{n(n-1)(n-2)(3n-5)}{24}.
\end{align}
From (\ref{13}), one has
\begin{equation}\label{ba}
\frac{B}{E}=\frac{1}{N}\sum_il_i-\frac{N-1}{2}.
\end{equation}

Defining
\begin{equation}
{\cal O}_2:=\left(\frac{\partial}{\partial\xi_1}\right)^2{\cal O},
\end{equation}
applying ${\cal O}_2$ on (\ref {lr}), and putting $\mathbf{s}=\mathbf{e}$,
one arrives at
\begin{align}\label{20}
\Bigg( \sum_i l_i^2 +\sum_{i<j}l_il_j \Bigg)\mathrm{van}
(l_1,\dots,l_N)=\mathrm{van}(l_1,\dots ,l_N)\Bigg[& p_{N+1}+q_{N+1}\frac{N\,B}{E}
\nonumber\\
&+\frac{N(N+1)\,C}{E}\Bigg],
\end{align}
from which
\begin{equation}\label{ca}
\frac{C}{E}=\frac{1}{N(N+1)}\left(\sum_il_i^2+\sum_{i<j}l_il_j\right)
-\frac{1}{2}\left(\sum_il_i\right)+\frac{(3N+2)(N-1)}{24}.
\end{equation}

Finally, defining
\begin{equation}
{\cal O}_3:=\frac{\partial^2}{\partial\xi_1\partial\xi_2}{\cal O},
\end{equation}
applying ${\cal O}_3$ on (\ref {lr}), and putting $\mathbf{s}=\mathbf{e}$,
one arrives at
\begin{align}
\Bigg( \sum_{i<j}l_il_j \Bigg)\mathrm{van}(l_1,\dots ,l_N)=
\mathrm{van}(l_1,\dots ,l_N)\Bigg[&q_N\,q_{N-1}- p_N+q_{N-1}\frac{N\,B}{E}\nonumber\\
&+\frac{N(N-1)(D-C)}{E}\Bigg],
\end{align}
which results in
\begin{equation}\label{da}
\frac{D}{E}=\frac{C}{E}+\frac{1}{N(N-1)}\Bigg(\sum_i l_i l_j\Bigg)-\frac{N-2}{2N}
\Bigg(\sum_i l_i\Bigg)+\frac{(N-2)(3N-5)}{24}.
\end{equation}

So, one has
\begin{equation}\label{26}
\ln\left[\frac{\chi_R(U)}{d_R}\right]=a\sum_i(s_i-1)+b\sum_i(s_i-1)^2+c\Bigg[\sum_i (s_i-1)\Bigg]^2+
\cdots,
\end{equation}
where
\begin{align}\label{abc}
a:=\frac{B}{E}=&\frac{1}{N}\sum_il_i-\frac{N-1}{2},\nonumber\\
b:=\frac{C}{E}-\frac{D}{2E}=&\frac{1}{2(N^2-1)}\Bigg(\sum_i l_i^2\Bigg)
-\frac{1}{2N(N^2-1)}\Bigg(\sum_i l_i\Bigg)^2\nonumber\\
&-\frac{1}{2N}\Bigg(\sum_i l_i\Bigg)+\frac{5N-6}{24},\nonumber\\
c:=\frac{1}{2}\left[\frac{D}{E}-\left(\frac{B}{E}\right)^2\right]=&
-\frac{1}{2N(N^2-1)}\Bigg(\sum_il_i^2\Bigg)+\frac{1}{2N^2(N^2-1)}
\Bigg(\sum_i l_i\Bigg)^2\nonumber\\
&+\frac{1}{24}.
\end{align}
\section{The large-$N$ limit of the U$(N)$ partition function}
In the large-$N$ limit, one introduces the continuous variables \cite{11}
\begin{align}
\phi (x)&:=-\frac{l(x)}{N},\nonumber\\
0\leq x&:=\frac{i}{N}\leq 1,
\end{align}
which represent the irreducible representation. (Note that
$l_i$ in this paper is the same as $n_i-i+N$ in \cite{11,20}). In the
large-$N$ limit,
\begin{equation}
\sum_i f(l_i)\to N\int_0^1\mathrm{d}x\;f[-N\phi (x)].
\end{equation}
So, in the large-$N$ limit,
\begin{align}\label{abc2}
a&=-N\left[ \int_0^1\mathrm{d}x\;\phi(x)+\frac{1}{2} \right],\nonumber\\
b&=\frac{N}{2}\left\{\int_0^1\mathrm{d}x\;\phi^2(x)
-\left[\int_0^1\mathrm{d}x\;\phi(x)\right]^2
+\int_0^1\mathrm{d}x\;\phi(x)+\frac{5}{12}\right\},\nonumber\\
c&=\frac{1}{2}\left\{-\int_0^1\mathrm{d}x\;\phi^2(x)
+\left[\int_0^1\mathrm{d}x\;\phi(x)\right]^2+\frac{1}{12} \right\}.
\end{align}
In the large-$N$ limit, the discrete eigenvalues
$s_j=e^{i\theta_j}$ are also represented by the eigenvalue density
function $\sigma (\theta )$ with $\theta \in [-\pi ,\pi ]$, and one has \cite{15}
\begin{equation}
\sum_i f(\theta_i)\to N\int_{-\pi}^{\pi}\mathrm{d}\theta\;\sigma(\theta)\,
f(\theta ).
\end{equation}
So, if $U\approx I$ (which means $\mathbf{s}\approx\mathbf{e}$) using
\begin{equation}
s_j-1=i\theta_j-\theta_j^2/2+\cdots,
\end{equation}
one arrives at
\begin{align}\label{34}
\sum_j(s_j-1)&\to N\left[ i\int\mathrm{d}\theta\;\sigma(\theta)\,
\theta-\frac{1}{2}\int\mathrm{d}\theta\;\sigma(\theta)\,\theta^2\right],\nonumber\\
\sum_j(s_j-1)^2&\to -N\int\mathrm{d}\theta\;\sigma(\theta)\,\theta^2,\nonumber\\
\left[\sum_j(s_j-1)\right]^2&\to
-N^2\left[\int\mathrm{d}\theta\;\sigma(\theta)\,\theta\right]^2.
\end{align}
Inserting (\ref{abc2}) and (\ref{34}) in (\ref{26}), one finds
\begin{align}\label{S1}
\ln\left[\frac{\chi_R(U)}{d_R}\right]=&-N^2\left[\int\mathrm{d}x\,\phi(x)+\frac{1}{2}\right]
\left[i\int\mathrm{d}\theta\;\sigma(\theta)\,\theta\right]\nonumber\\
&-\frac{N^2}{2}\left\{\int\mathrm{d}x\;\phi^2(x)-\left[\int\mathrm{d}x\;\phi(x)\right]^2
-\frac{1}{12}\right\}\nonumber\\
&\times\left\{\int\mathrm{d}\theta\;\sigma(\theta)\,\theta^2
-\left[\int\mathrm{d}\theta\;\sigma(\theta)\,\theta\right]^2\right\} .
\end{align}
Defining
\begin{align}\label{V}
Q(U)&:=\int\mathrm{d}\theta\;\sigma(\theta)\,\theta,\nonumber\\
V(U)&:=\int\mathrm{d}\theta\;\sigma(\theta)\,\theta^2
-\left[\int\mathrm{d}\theta\;\sigma(\theta)\,\theta\right]^2,
\end{align}
it is seen that under the translation $\theta\to\theta+\alpha$,
$Q\rightarrow Q+\alpha,$ while $V$ remains invariant. The translation
$\theta\to\theta+\alpha$ does not change the SU$(N)$ factor of $U$, but it
does change the U$(1)$ part of $U$. So, the logarithm of the character is
the sum of two terms, one comming from the SU$(N)$ part, the other from
the U$(1)$ part, as was expected from (\ref{5}).

For the remaining part of the partition function one has
\begin{equation}
d_R^\eta\, e^{-\frac{A}{2N}C_{2\,R}}=:e^{S_0},\qquad\eta:=2-2g,
\end{equation}
and (following \cite{11})
\begin{equation}\label{S0}
S_0[\phi]=-\frac{N^2\,A}{2}\int_0^1\mathrm{d}x\;\left[\phi(x)+\frac{1}{2}\right]^2
+\frac{N^2\,\eta}{2}\int\mathrm{d}x\;\mathrm{d}y\;\log|\phi(x)-\phi(y)|+{\cal C},
\end{equation}
where ${\cal C}$ is a constant. The large-$N$ limit of the
partition function (\ref{z}) then becomes the following functional
integral
\begin{equation}\label{39}
Z_{g,n}(U_1,\cdots,U_n;A)=\int{\cal D}\phi\;e^{S[\phi]},
\end{equation}
where
\begin{equation}
S[\phi ]=S_0[\phi]+S'[\phi],
\end{equation}
in which
\begin{align}\label{41}
S'[\phi]=&-iN^2\,Q\left[\int\mathrm{d}x\;\phi(x)+\frac{1}{2}\right]\nonumber\\
&-\frac{N^2}{2}\,V\left\{\int\mathrm{d}x\;\phi^2(x)-
\left[\int\mathrm{d}x\;\phi(x)\right]^2-\frac{1}{12}\right\},
\end{align}
where
\begin{align}\label{42}
Q&=\sum_j Q(U_j),\nonumber\\
V&=\sum_j V(U_j).
\end{align}
Defining
\begin{align}
q&:=\int_0^1\mathrm{d}x\;\left[\phi(x)+\frac{1}{2}\right],\nonumber\\
\psi(x)&:=\phi(x)+\frac{1}{2}-q,
\end{align}
one arrives at
\begin{equation}
Z_{g,n}(U_1,\cdots,U_n;A)=Z_1\,Z_2,
\end{equation}
where
\begin{align}
Z_1:=&{\cal N}\int\mathrm{d}q\;\exp\left[-N^2\left(\frac{A\,q^2}{2}+iQ\,q\right)\right],\nonumber\\
=&\exp\left(-\frac{N^2\,Q^2}{2A}\right),
\end{align}
and
\begin{align}\label{Z}
Z_2:=e^{N^2V/24}\int{\cal D}\psi\;\exp\Bigg\{&-\frac{N^2}{2}
\Bigg[(A+V)\int\mathrm{d}x\;\psi^2(x)\nonumber\\
&-\eta\int\mathrm{d}x\;\mathrm{d}y\;\log|\psi(x)-\psi(y)|\Bigg]\Bigg\}.
\end{align}
${\cal N}$ is a normalization constant. It is seen that $Z_2$ is in fact
equal to the partition function on a genus-$g$ surface without boundaries,
with the surface area equal to $A+V$:
\begin{equation}
Z_2=Z_{g,0}(A+V).
\end{equation}
So,
\be\label{45}
\log[Z_{g,n}(U_1,\cdots,U_n;A)]=\frac{N^2}{2}\left(\frac{V}{12}-\frac{Q^2}{A}\right)
+\log[Z_{g,0}(A+V)].
\end{equation}
As the first term is a smooth function of $A$, the phase
transition comes from the second term, which is known to be trivial for
$g>0$ \cite{11}. So there is a phase transition only for $\Sigma_{0,n}$,
and the transition occures at
\begin{equation}
A_{\mathrm{c}}=\pi^2-V.
\end{equation}
The boundary conditions do not change the structure of the phase transition:
it is still a third order phase transition. They do, however, change the
critical area.

As an example, let us consider a sphere with one hole, that is a
disc. The critical area of a disc with a boundary condition such that the
eigenvalue density $\sigma(\theta)$ is even, has
been found in \cite{15}:
\begin{equation}\label{Ac}
A_{\mathrm{c}}=\pi\left[\int\frac{\mathrm{d}\theta\;\sigma(\theta)}
{\pi-\theta}\right]^{-1}
\end{equation}
For $U\approx I$, $\sigma(\theta)$ is nonnegligible only at $\theta$ near zero.
Expanding the denominator of (\ref{Ac}) up to second order in $\theta$, one
arrives at
\begin{align}
A_{\mathrm{c}}=&\pi^2\left[\int\mathrm{d}\theta\;\sigma(\theta)
+\frac{1}{\pi}\int\mathrm{d}\theta\;\sigma(\theta)\,\theta
+\frac{1}{\pi^2}\int\mathrm{d}\theta\;\sigma(\theta)\,\theta^2+\cdots
\right]^{-1},\nonumber\\
=&\pi^2\left[1+\frac{1}{\pi^2}\int\mathrm{d}\theta\;\sigma(\theta)\,\theta^2+\cdots
\right]^{-1},\nonumber\\
=&\pi^2-\int\mathrm{d}\theta\;\sigma(\theta)\,\theta^2+\cdots.
\end{align}
This is consistent with our general result, since in this case the second term of
$V$ in (\ref{V}) vanishes.
\section{The partition function for other groups}
The character expression introduced in eq.(\ref{ch}) is for
the group $U(N)$. It can be used, however, for SU$(N)$ as well. If
$U\in\mathrm{SU}(N)$, then the product of the eigenvalues of $U$ is equal to one,
and it is easily seen that in this case, translating all $l_j$'s by a fixed integer
does not change the character. In fact, for SU$(N)$ one can use the same $l_1$ to
$l_{N-1}$ to characterize the representation, or use the same results obtained for
U$(N)$ but with $Q=0$. So, one arrives at
\begin{equation}\label{sim}
Z_{g,n}(U_1,\dots,U_n;A)=Z_{g,0}(A+V).
\end{equation}

For gauge groups other than U$(N)$ and SU$(N)$, another approach is followed.
Suppose that the gauge group is simple. A group element is characterized by
$D$ parameters $x^\alpha$ like
\begin{equation}
U=\exp(x^\alpha\,T_\alpha),
\end{equation}
where $T_\alpha$'s are the generators of the group. If $U\approx I$, then
\begin{equation}\label{48}
U=1+x^\alpha\,T_\alpha+\frac{1}{2}x^\alpha\,x^\beta\,T_\alpha\,T_\beta+\cdots,
\end{equation}
from which,
\begin{equation}\label{49}
\chi_R(U)=d_R+\frac{1}{2}x^\alpha\,x^\beta\,\chi_R(T_\alpha\,T_\beta)+\cdots,
\end{equation}
in which use has been made of the fact that the representations of the
generators of simple groups are traceless. For a simple group,
\begin{equation}
\chi_R(T_\alpha\,T_\beta)=\frac{d_R\, C_{2\,R}}{d_G}\,\Omega_{\alpha\beta},
\end{equation}
where $\Omega$ is the Killing form of the group, and $d_G$ is the dimension of the group.
So, up to second order in $x^\alpha$'s,
\begin{equation}
\frac{\chi_R}{d_R}=\exp \left(\frac{C_{2\,R}}{2d_G}\,\Omega_{\alpha\beta}\,
x^\alpha\,x^\beta\right).
\end{equation}
So from (\ref{z}), one arrives at (\ref{sim}) with
\begin{equation}
V=-\frac{N}{d_G}\,\Omega_{\alpha\beta}\sum_j x_j^\alpha\,x_j^\beta,
\end{equation}
where the summation is over the boundaries.

\end{document}